\documentclass[final,1p,times]{elsarticle}
\usepackage{amssymb}
\usepackage{amsmath}
\usepackage{amsthm}

\makeatletter
\@ifundefined{textcolor}{}
{
 \definecolor{BLACK}{gray}{0}
 \definecolor{WHITE}{gray}{1}
 \definecolor{RED}{rgb}{1,0,0}
 \definecolor{GREEN}{rgb}{0,1,0}
 \definecolor{BLUE}{rgb}{0,0,1}
 \definecolor{CYAN}{cmyk}{1,0,0,0}
 \definecolor{MAGENTA}{cmyk}{0,1,0,0}
 \definecolor{YELLOW}{cmyk}{0,0,1,0}
}

\@ifundefined{date}{}{\date{15-Jan-2022}}
\@ifundefined{textcolor}{}
{
 \definecolor{BLACK}{gray}{0}
 \definecolor{WHITE}{gray}{1}
 \definecolor{RED}{rgb}{1,0,0}
 \definecolor{GREEN}{rgb}{0,1,0}
 \definecolor{BLUE}{rgb}{0,0,1}
 \definecolor{CYAN}{cmyk}{1,0,0,0}
\definecolor{MAGENTA}{cmyk}{0,1,0,0}
 \definecolor{YELLOW}{cmyk}{0,0,1,0}
}

\@ifundefined{date}{}
{\date{Apr-13- 2023}}

\makeatother

\begin{document}
\begin{frontmatter}
\title{Inflation and Fractional Quantum Cosmology}

\author[1,2]{S. M. M. Rasouli}
\author[3]{Emanuel W. de Oliveira Costa}
\author[1]{Paulo Vargas Moniz}
\author[3]{Shahram Jalalzadeh}

\address[1]{Departamento de F\'{i}sica,
Centro de Matem\'{a}tica e Aplica\c{c}\~{o}es (CMA-UBI),
Universidade da Beira Interior,
 Rua Marqu\^{e}s d'Avila
e Bolama, 6200-001 Covilh\~{a}, Portugal.}

\address[2]{Department of Physics, Qazvin Branch, Islamic Azad University, Qazvin, Iran}
\address[3]{Departamento de F\'{i}sica, Universidade Federal de Pernambuco,
Recife, PE 50670-901, Brazil}


\begin{abstract}
The Wheeler--DeWitt equation for a flat and compact Friedmann--Lema\^{i}tre--Robertson--Walker cosmology at the pre-inflation epoch is studied in the contexts of the standard and fractional quantum cosmology. Working within the semiclassical regime and applying the WKB approximation,  we show that some fascinating  consequences are obtained for our simple fractional scenario that are completely different from their corresponding standard counterparts: (i) The conventional de Sitter behavior of the inflationary universe for constant potential is replaced by a power-law inflation. (ii) The non-locality of the Riesz's fractional derivative produces a power-law inflation that depends on the fractal dimension of the compact spatial section of space-time, independent of the energy scale of the inflaton.
 \end{abstract}

\medskip

\end{frontmatter}

\section{Introduction}
\label{SecI}

Inflation is widely accepted as a formal solution to cosmological problems such as horizon and flatness~\cite{Guth:1980zm,Albrecht:1982wi,Linde:1981mu}. The conventional approach to controlling inflation is that a scalar field with an appropriate potential, such as the Coleman--Weinberg potential~\cite{Coleman:1973jx}, dominates the energy density of the universe from the start. In early theories of inflation, this energy density promotes a rapid expansion for the scale factor of the universe~\cite{Steinhardt:1984jj,Ellis:1982ws}. This accelerated expansion would be exponential, similar to de Sitter space, or power law, according to which the scalar field gradually decreases to the global minimum of its~potential.

In addition to the well-known standard models of inflation, recent scenarios using non-commutative cosmological models, Generalized Uncertainty Principle (GUP) and other modified gravity models have shown that  some problems with the standard models (such as graceful exit and the Hubble constant problem) can be solved, see, for instance~\cite{Rasouli:2014dba,Rasouli:2018lny,Aghababaei:2021gxe,Oliveira-Neto:2021zqy,Maldonado:2021aze,Rasouli:2022hnp}.

It should be noted that quantum cosmology is a suitable method to study the essential initial conditions for the emergence of an inflationary phase. Tryon proposed in 1973 that a closed universe~\cite{Tryon:1973xi} could emerge spontaneously as a quantum fluctuation. He realized that in a spatially closed universe, all conserved charges are zero. Consequently, no conservation law prevents such a universe from forming spontaneously.
Furthermore, according to general relativity, at the instant of creation of such a universe, not only matter fields, but also space-time itself is created, and there has not been anything earlier than that. In fact, if we assume that the universe is spatially homogeneous, isotropic, compact, and simply compact, the only feasible choice to have a space without a boundary is a closed space.
The volume of open and flat spaces that are simply connected is infinite. By removing the restriction of simple connectivity, more sophisticated yet finite volume spaces for all possible flat, open, and close homogeneous and isotropic spaces can be obtained (please see~\cite{Jalalzadeh:2020bqu} and references therein). In this work, we assume that the cosmic manifold is spatially compact since we are trying to understand the universe as a whole and it is implicitly inconceivable that the universe has a spatial boundary. This implies that the three-dimensional spacelike hypersurfaces are compact (in mathematical language, a compact manifold without a boundary is called a closed manifold).

Intuitively, a universe formed by a quantum fluctuation should be exceedingly small, having a Planck length linear size. This was initially a major challenge since it was unclear how to create the enormous universe we live in from a tiny compact quantum cosmological model. With the emergence of inflationary theories in which the universe goes through a de Sitter phase of exponential growth, the dilemma has vanished. All scales in the universe are expanded by a gigantic factor $\exp(Ht)$ as a result of inflation, where $H$ is the constant expansion rate and $t$ is the duration of the inflationary phase or $t^\alpha$ where $\alpha$ must be greater than one and ideally at least of the order of ten.

Given the centrality of inflation in our present understanding of cosmology, it is reasonable and vital to try to grasp its details within the context of fractional quantum cosmology. The history of fractional derivatives is as long as that of classical calculus. A fractional derivative is a generalization of the integer-order derivative. It originated in the letter regarding the meaning of half-order derivative from L' H\^opital to Leibnitz in 1695 and is a promising tool for explaining various phenomena. Various definitions of fractional derivatives exist in the literature, including Riesz, Riemann--Liouville, Caputo, Hadamard, Marchand, and Griinwald--Letnikov, among others~\cite{book1:2006,book2:1999}. Recent quantum gravity conclusions have provided an essential push for the increased use of fractional calculus in quantum theory. Various approaches to quantum gravity, such as asymptotically safe quantum gravity~\cite{Lauscher:2005qz,2006LRR5N,Reuter:2011ah,Reuter:2012xf}, causal dynamical triangulations~\cite{Ambjorn:2005db,Modesto:2009qc,Ambjorn:2012jv,Sotiriou:2011mu}, loop quantum gravity, and spin foams~\cite{Modesto:2008jz,Perez:2012wv,Rovelli:2004tv}, Ho\v{r}ava--Lifshitz gravity~\cite{Horava:2009if,Sotiriou:2010wn}, non-local quantum gravity~\cite{Biswas:2011ar,Calcagni:2014vxa,Modesto:2011kw}, and others, all lead to the same result: the dimension of space-time changes with scale. As a result of the abnormal scaling of the space-time dimension, all known theories of quantum gravity are multiscale.
 There are numerous applications for fractional calculus and active research in this field. Fractional quantum mechanics has been used to model fractional space-time in gravity and cosmology~\cite{Jalalzadeh:2022uhl,Rasouli:2021lgy,Moniz:2020emn,Calcagni:2021aap,Calcagni:2012cn,Calcagni:2016xtk,Calcagni:2021mmj,Garcia-Aspeitia:2022uxz}, and the fractional quantum field theory~\cite{tarasov2014fractional,Calcagni:2021ljs,2006PhyA269L}.

The main objective of this work is to reexamine the standard inflation in the context of  fractional quantum cosmology (FQC) as a novel underlying scenario for studying the early state of the universe.
Concretely, this study aims to employ the FQC with a minimally coupled scalar field and a compact 3-space as a probe model. We are aware of the limited breadth. However, fascinating aspects can be found. In the next section, we present a review of the model in the context of the canonical quantum cosmology and obtain a semi-classical wavefunction.
Section \ref{FQC} focuses on the fractional counterpart of the model mentioned above. Then, by analyzing the implications of fractional quantum cosmology, we examine why they may be more intriguing than those reached in Section \ref{SecII}.


\section{Wheeler-DeWitt Equation in Slow Roll Regime}
\label{SecII}
\indent
Let us consider the Friedmann--Lema\^{i}tre--Robertson--Walker (FLRW) metric
\begin{eqnarray}
\label{FRW-met-1}
{\rm d}s^2 = -N^2(t){\rm d}t^2
+a^2(t)\biggl(\frac{{\rm d}r^2}{1-\mathcal Kr^2}+
r^2({\rm d}\theta ^2+\sin ^2\theta {\rm d}\varphi ^2)\biggr),
\end{eqnarray}
where $N(t)$ is the lapse function, $a(t)$ is the scale factor, and we will assume that the space-time manifold, $\mathcal M$, is a spatially compact and globally
hyperbolic Lorentzian manifold. We claimed that $\mathcal M$ is spatially closed since we want to investigate the cosmological model as a whole. It is basically impossible to imagine that the universe has a spatial boundary. Consequently,
the spacelike sections of the space-time manifold, $(\Sigma,h)$, have a finite volume with the
normalized curvature index $\mathcal K$ taking values $0, \pm1$.
Different ranges of variation for the
coordinates $(r, \theta, \varphi )$ determine different topologies.
Let us rewrite the metric as
\begin{eqnarray}
\label{gen-met}
{\rm d}s^2=g_{\mu\nu}{\rm d}x^\mu{\rm d}x^\nu=-N^2(t){\rm d}t^2+
a^2(t)h_{ij}{\rm d}x^i{\rm d}x^j.
\end{eqnarray}
Then, we can define the volume of the spacelike hypersurfaces as
\begin{equation}
\label{defvol}
{\mathcal V}_{{\mathcal K}}\equiv \int {\rm d}^3x\sqrt{h}.
\end{equation}
Therefore, scalar
curvature of the spacelike hypersurface $(\Sigma, h)$ of the above
metric is constant and equal to $6\mathcal K/a^2$. One can show that any compact Riemannian
three-manifold with constant curvature is homeomorphic to
$\tilde\Sigma/\Gamma$, where $\Gamma$ denotes the group of the covering transformations and $\tilde\Sigma$ is the universal covering space, which is either $\mathbb R^3$ (3-dimensional Euclidean space), $\mathbb S^3$ (3-sphere), or
$\mathbb H^3$ (3-dimensional hyperbolic space) regarding the sign of $\mathcal K$ ($\mathcal K=0$, $\mathcal K=1$, or $\mathcal K=-1$, respectively). Therefore, in
three dimensions, an oriented compact and without boundary  space is now a polyhedron
whose faces are identified in pairs. For an overview, see
Ref.~\cite{Jalalzadeh:2020bqu}.

As we would like to investigate a new scenario regarding
an inflationary universe, let us consider the same action
corresponding to standard inflationary theories:
\begin{equation}
\label{act-inf}
S=\int {\rm d}^4x\sqrt{-g}\left[\frac{R}{16\pi G}-\frac{1}{2}g^{\mu\nu}{\nabla}_\mu\phi{\nabla}_\nu\phi
-V(\phi)\right],
\end{equation}
where $\phi$ is a scalar field minimally coupled to the
Ricci scalar $R$, and we used the units where $c=1=\hbar$.
As during inflation only, the inflaton $\phi$ dominates the dynamics, the
action does not contain any other matter fields.

By substituting the Ricci scalar corresponding to the metric \eqref{FRW-met-1}, i.e.,
\begin{equation}
\label{FRW-Ricci}
R=\frac{6}{N^2}\left[\frac{\ddot{a}}{a}+\left(\frac{\dot{a}}{a}\right)^2-\left(\frac{\dot{N}}{N}\right)\left(\frac{\dot{a}}{a}\right)
+\mathcal K\left(\frac{N}{a}\right)^2\right],
\end{equation}
into action functional \eqref{act-inf}, we obtain Arnowitt--Deser--Misner (ADM) action of the model
\begin{equation}
\label{FRW-action}
S_\text{ADM}=\frac{3{\mathcal V}_{{\mathcal K}}}{8\pi G}\int {\rm d}t Na^3\left[\frac{\mathcal K}{a^2}
-\frac{1}{N^2}\left(\frac{\dot{a}}{a}\right)^2\right]-{{\mathcal V}_{{\mathcal K}}}\int {\rm d}tNa^3\left[-\frac{1}{2N^2}\dot{\phi }^2
+V(\phi )\right],
\end{equation}
where an overdot denotes differentiation with respect to $t$.

The conjugate momenta associated with $a$ and $\phi$, respectively, are given by
\begin{equation}
\label{momenta}
\Pi _a = -\frac{3{\mathcal V}_{{\mathcal K}}}{4\pi G}\frac{a\dot{a}}{N}, \quad
\Pi _{\phi } = \frac{a^3{\mathcal V}_{{\mathcal K}}}{N}\dot{\phi },
\end{equation}
by which we can easily obtain the canonical ADM Hamiltonian:
\begin{eqnarray}
\label{Hamil1}
H_\text{ADM}= N\biggl[-\frac{2\pi G}{3{\mathcal V}_{{\mathcal K}}a}{\Pi ^2_a}+\frac{1}{{\mathcal V}_{{\mathcal K}}}\frac{\Pi ^2_{\phi }}{2a^3}
-\frac{3{\mathcal K}{\mathcal V}_{{\mathcal K}}}{8\pi G}a +{{\mathcal V}_{{\mathcal K}}}a^3V(\phi )\biggr].
\end{eqnarray}
The above ADM Hamiltonian leads to the super-Hamiltonian constraint
\begin{equation}\label{super}
 \mathcal H=   -\frac{2\pi G}{3{\mathcal V}_{{\mathcal K}}a}{\Pi ^2_a}+\frac{1}{{\mathcal V}_{{\mathcal K}}}\frac{\Pi ^2_{\phi }}{2a^3}
-\frac{3{\mathcal K}{\mathcal V}_{{\mathcal K}}}{8\pi G}a +{{\mathcal V}_{{\mathcal K}}}a^3V(\phi )=0.
\end{equation}

To obtain the Wheeler--DeWitt (WDW) equation, we apply the quantization map
\begin{equation}
\label{FO}
\Pi _a^2\rightarrow -a^{-p}\frac{\partial }{\partial a}
(a^p\frac{\partial }{\partial a}), \quad
\Pi _{\phi }^2 \rightarrow -\frac{{\partial }^2}{\partial \phi ^2},
\end{equation}
where the ordering parameter $p$ is considered for the factor-ordering ambiguity~\cite{Vilenkin:1987kf}.
By the action of the super-Hamiltonian operator $\hat{\mathcal H}$ on a wave function
$\Psi (a,\phi )$, we can therefore write the WDW equation:
\begin{equation}
\label{wdw1}
\frac{{ \partial^2\Psi}}{{\rm \partial}a^2}+
\frac{p}{a}\frac{{ \partial\Psi}}{{ \partial}a}-
\frac{3m_\text{P}^2}{4\pi  a^2}\frac{{\partial^2\Psi}}{{ \partial}\phi ^2}
-\frac{9 {\mathcal V}_{{\mathcal K}}^2 m_\text{P}^4}{16 \pi^2}a^2
\left[{\mathcal K}-\frac{8\pi}{3m_\text{P}^2}{a}^2 V(\phi)\right]\Psi=0,
\end{equation}
where $m_\text{P}=1/\sqrt{G}$ is the Planck mass.

As mentioned, in this work we are interested in
investigating a fractional counterpart of an inflationary model established during a {\it `slow roll'} regime. For such a  special but familiar scenario, we can ignore the $\phi$--dependence of the wave function. Furthermore, in such a regime, we can assume that the scalar potential plays the role of an effective cosmological constant~\cite{Linde:1990xn}. More precisely, we define $ \Lambda\equiv (8\pi V_0)/m_\text{P}^2={\rm constant}$. Additionally, let us define $L\equiv\sqrt{3/\Lambda}$, and assume $p=0$. In the following, we are interested in semi-classical approximation, and in this case the factor ordering does not play a crucial role. According to Linde~\cite{Linde:2004nz}, a compact flat universe, such as the toroidal universe ($\mathcal K=0$, $\Sigma=\mathbb T^3=\mathbb S^1\times \mathbb S^1\times \mathbb S^1$), is the simplest case to solve the problem of initial conditions for low-scale inflation. While the quantum creation of a closed or infinite open inflationary universe is exponentially inhibited, the toroidal universe is not.

Consequently, substituting $p=0$ and $\mathcal K=0$ into WDW Equation \eqref{wdw1}, we obtain
\begin{eqnarray}
\label{wdw1a}
\frac{{\rm d}^2}{{\rm d}a^2}\Psi (a)+
\frac{9 {\mathcal V}_{{0}}^2 m_\text{P}^4}{16 \pi^2L^2}a^4\Psi (a)=0.
\end{eqnarray}
A solution for \eqref{wdw1a} in terms of the Bessel function is:
\begin{eqnarray}
\label{Sol1}
\Psi (a)=\mathcal N\sqrt{a}J_\frac{1}{6}\left(\frac{\mathcal V_0m_P^2}{4\pi L}a^3\right),
\end{eqnarray}
where $\mathcal N$ is a normalization constant. The asymptotic form of the Bessel function, for large values of the scale factor, i.e.,
\begin{equation}\label{sh3}
    \frac{\mathcal V_0a^3}{4\pi l_P^2 L}\gg1,
\end{equation}
yields the asymptotic form of the wave function
\begin{equation}\label{sh1}
    \Psi(a)=\mathcal N\cos\left(\frac{\mathcal V_0m_P^2}{4\pi L}a^3
    -\frac{1}{3}\right).
\end{equation}
This wavefunction collects both the
expanding $\exp(-i\frac{\mathcal V_0m_P^2}{4\pi L}a^3)$ and contracting $\exp(i\frac{\mathcal V_0m_P^2}{4\pi L}a^3)$ phases of the cosmological model simultaneously.

We now discuss the classical limit of the wavefunction. First, we substitute the wavefunction written in form ({{Equation \eqref{SS} is the single component WKB solutions to the WDW
equation in the oscillatory regime~\cite{Cooke2010AnIT}. To study the oscillatory behavior of the particular differential equations, see~\cite{el2020new}.}})  

\begin{equation}
    \label{SS}
    \Psi(a)=\exp(-iS),
\end{equation}
 into the WDW Equation (\ref{wdw1a}).
Straightforward differentiation plus condition (\ref{sh3}) (which is equivalent to WKB) ({{The semi-classical approximation and also the WKB wave function are important concepts in the context of quantum cosmology. It has been believed that the semi-classical approximation is employed to avoid the ambiguities arising from operator ordering issues in the Wheeler--DeWitt equation, as well as problems in the path integral formulation of the wave function. The WKB wave function is best suited to approximate the wave function in the semi-classical regime~\cite{Cooke2010AnIT}}}), i.e., $d^2S/da^2\ll (dS/da)^2$) lead to
\begin{equation}
    \label{sh4}
    -\Big|\frac{{\rm d}S}{{\rm d}a}\Big|^2+
\frac{9 {\mathcal V}_{{0}}^2 m_\text{P}^4}{16 \pi^2L^2}a^4=0.
\end{equation}
We recognize that \eqref{sh4} is the Hamilton--Jacobi equation in classical cosmology, which is equivalent to the Hamiltonian constraint. As we know, the momentum in the classical Hamilton--Jacobi theory is given by $\Pi_a=-dS/da$. Therefore, with this definition and the corresponding relation in (\ref{momenta}), we obtain $\dot a/a=1/L$ or $a(t)=a(t_0)\exp((t-t_0)/L)$.

Thus, we have shown that the scale factor grows exponentially during the inflationary epoch, $t_i\leq t\leq t_f$, which is assumed to occur in a classical region,
\begin{equation}\label{sh00}
    \frac{a(t_f)}{a(t_i)}=\exp{\left(\frac{t_f-t_i}{L}\right)}.
\end{equation}
This solution implies the expansion rate during the de Sitter expansion. The so-called number of e-folds $N_e$ is often introduced as $N_e=\ln(a(t_f)/a(t_i))$. If we assume that the inflation occurred during a very small time interval $t_f-t_i=10^{-37}~s=10^8t_P$, the minimum number of e-foldings necessary for resolving the
standard problems of the Big Bang is $N_e\simeq 60$. Substituting these values into the above de Sitter solution, we obtain $L\simeq 10^7 l_p$.

\section{Fractional Quantum Cosmology for a Slow Roll Regime}
\label{FQC}
In this section, we investigate our model in the context of the FQC.
In this context, we first provide a brief overview of the FQC framework.
In order to construct a fractional WDW equation, it is best to obtain the fractional Schr\"odinger equation (SE).

The classical Hamiltonian of a particle with mass $m$, space coordinates $\mathbf{r}$, and momentum $\mathbf{p}$ is $H=\frac{\mathbf{p}^2}{2m}+V(\mathbf{r})$.
{Focusing on the quadratic dependence of $\mathbf{p}$ in the above equation might inspire to adopt a different function for the kinematic term so that it does not conflict with basic principles of mechanics. An acceptable approach from theoretical physics for this motivation is the Feynman path integral approach to quantum mechanics~\cite{feynman2010quantum}, first developed by Laskin~\cite{Laskin:1999tf}.
Non-relativistic quantum mechanics was reconstructed by Feynmann and Hibbs as a path integral over the Brownian paths. Moreover, the space fractional quantum mechanics formulated by Laskin applying new fractional path integral on the basis of the L\'{e}vy flight~\cite{Laskin:1999tf,Laskin:2002zz}. In addition, Naber obtained the time fractional SE on the basis of fractional Brownian motion~\cite{Naber2004TimeFS}, see also~\cite{Achar2013TimeFS}. Inspired by the above mentioned works, the generalized fractional SE was obtained where both the space and time derivatives were replaced by fractional counterparts.}

{In what follows, let us present a brief overview of the
fractional quantum dynamics, which is generated by the Hamiltonian function $H_\alpha(\mathbf p, \mathbf r) := D_\alpha |\mathbf p|^\alpha+ V(\mathbf r)$ (for a detailed review, see~\cite{Laskin:2010ry}),}
where $D_\alpha$ is a scale coefficient, and the L\'evy's fractional parameter $\alpha$ relevant to the concept of L\'evy path~\cite{par1} is set to $1< \alpha\leq 2$.
Then, in order to construct the corresponding quantum Hamiltonian of the system,  we should apply the standard canonical quantization procedure, i.e., $(\mathbf{r},\mathbf{p})\rightarrow (\mathbf{r},\mathbf{-i\hbar\nabla})$. Hence, we obtain $H_\alpha= D_\alpha
(-\hbar^2 \Delta)^{\alpha/2}+V(\mathbf{r})$.
Now  the usual SE,
\begin{equation}
 i\hbar \frac{\partial
 \psi(\mathbf{r},t)}{\partial  t}=
-\frac{\hbar^2}{2m}\Delta \psi(\mathbf{r},t)
+ V(\mathbf{r},t) \psi(\mathbf{r},t),
\label{7-0a}
\end{equation}
 can be extended to its fractional counterpart by replacing the ordinary Laplace operator $\Delta$ with the Riesz fraction $(-\hbar^2 \Delta)^{\alpha/2}$:
\begin{equation}
    -\frac{\hbar^2}{2m}\Delta\longrightarrow D_\alpha
(-\hbar^2 \Delta)^{\alpha/2}.
\end{equation}

Corresponding to a three-dimensional
Euclidean space, the Riesz fractional Laplacian
$(-\hbar^2 \Delta)^{\alpha/2}$ in terms of the
Fourier transformation is written as~\cite{Rie}
\begin{equation}
 \begin{array}{cc}
(-\hbar^2 \Delta)^{\alpha/2} \psi(\mathbf{r},t)
=\mathcal F^{-1}|\mathbf{p}|^\alpha\mathcal F\psi(\mathbf{r},t)\\
\\
=\displaystyle\frac{1}{(2\pi\hbar)^3}
\int d^3 p e^{i\frac{\mathbf{p}\cdot \mathbf{r}}{\hbar}}
|\mathbf{p}|^\alpha
\int e^{-i\frac{\mathbf{p}\cdot \mathbf{r}'}{\hbar}}\psi(\mathbf{r}',t)d^3r',
\end{array}
\end{equation}
where $\mathbf{p}=\sqrt{p_1^2+p_2^2+p_3^2}$.
Finally, the fractional SE is proposed as
\begin{equation}
    i\hbar \frac{\partial \psi(\mathbf{r},t)}{\partial  t}=
D_\alpha
(-\hbar^2 \Delta)^{\alpha/2} \psi(\mathbf{r},t)
+ V(\mathbf{r},t) \psi(\mathbf{r},t).
\label{7-0}
\end{equation}

Now, let us focus on our model. As for a homogeneous
cosmological model, in addition to the supermomentum
constraint for $H_i$, the shift function $N_i$,
without neglecting any of Einstein equations, can be set to
zero, the superspace WDW equation therefore reduces to the minisuperspace WDW equation:
\begin{equation}\label{WDW}
    \left\{\frac{1}{2}\Box+U(q^\nu)\right\}\Psi(q^\nu)=0,~~~\nu=0,\dots,N-1.
\end{equation}
In Equation \eqref{WDW}, $\Box=\frac{1}{\sqrt{-f}}
\partial_\alpha(\sqrt{-f}f^{\alpha\beta}\partial_\beta)$ is the d'Alembertian operator, $q^\alpha$ and $U(q^\nu)$ are coordinates of $N$-dimensional minisuperspace and potential, respectively;
$f_{\alpha\beta}$ stands for the corresponding minisuperspace metric
with signature $(-,+,+,\dots,+)$.

Regarding obtaining the fractional counterpart of the
WDW Equation \eqref{WDW}, the usual d'Alembertian operator should be replaced by~\cite{Rie,Tarasov:2018zjg,Calcagni:2021aap,Calcagni:2011kn}
\begin{equation}
(-\Box)^{\alpha/2} \Psi(q^\alpha)
=\mathcal F^{-1}(|\mathbf{p}|^\alpha(\mathcal F\Psi(\mathbf p)),
\end{equation}
where $|\mathbf p|=\sqrt{-p_0^2+p_ip^i}$, $i=1,2,\dots,N$, and $\mathcal F$ denotes Fourier transformation. Hence, the fractional counterpart of the WDW Equation (\ref{WDW}) will be~\cite{Jalalzadeh:2020bqu,Moniz:2020emn,Rasouli:2021lgy,Jalalzadeh:2021gtq,Jalalzadeh:2022uhl}
\begin{equation}\label{WDWsh}
    \left\{\frac{m_P^{2-\alpha}}{2}(-\Box)^\frac{\alpha}{2}-U(q^\nu)\right\}\Psi(q^\nu)=0,~~~\nu=0,\dots,N-1.
\end{equation}

As seen, the WDW Equation (\ref{wdw1a}) is actually a one-dimensional SE with zero energy.
Hence the final step of our procedure here is to obtain a particular
fractional WDW equation similar to that proposed above for the SE case.
Consequently, we can replace the ordinary derivative with the fractional Riesz derivative~\cite{Jalalzadeh:2020bqu,Moniz:2020emn,Rasouli:2021lgy,Jalalzadeh:2021gtq,Jalalzadeh:2022uhl}:
\begin{equation}
    \label{2-8}
    -\frac{d^2}{{\rm d}a^2}\longrightarrow m_\text{P}^{2-\alpha}
    (-\frac{{\rm d}^2}{{\rm d}a^2})^\frac{\alpha}{2},~~~~1<\alpha\leq2.
\end{equation}
Therefore, we obtain the fractional version of the WDW Equation (\ref{wdw1a}) for our herein model~as
\begin{eqnarray}
\label{FWDW}
-\left(-\frac{{\rm d}^2}{{\rm d}a^2}\right)^\frac{\alpha}{2}\Psi (a)+
\frac{9 {\mathcal V}_{0}^2 m_\text{P}^{\alpha+2}}{16 \pi^2L^2}a^4\Psi (a)=0.
\end{eqnarray}
Note that in a particular case where $\alpha=2$, Equation \eqref{FWDW} reduces to its standard counterpart \eqref{wdw1a}.
To obtain the semi-classical solution of this fractional WDW equation, we use the alternative equivalent representation of the Riesz fractional derivative (or fractional Laplacian)~\cite{pozrikidis2018fractional}, given by
\begin{equation}
    \label{sh5}
    -\left(-\frac{{\rm d}^2}{{\rm d}a^2}\right)^\frac{\alpha}{2}\Psi(a)=c_{1,\alpha}\int_0^\infty\frac{\Psi(a-v)-2\Psi(a)+\Psi(a+v)}{v^{\alpha+1}}{\rm d}v,
\end{equation}
where $c_{1,\alpha}=\frac{\alpha 2^{\alpha-1}}{\sqrt{\pi}}\frac{\Gamma((1+\alpha)/2)}{\Gamma((2-\alpha)/2)}$. Inserting the exponential form of the wavefunction, (\ref{SS}), into the above form of the fractional Laplacian,
expanding $\Psi(a\pm v)$ in Taylor series about $a$, and using semi-classical approximation, we obtain
\begin{equation}
    \label{sh6}
    -\left(-\frac{{\rm d}^2}{{\rm d}a^2}\right)^\frac{\alpha}{2}\Psi(a)=c_{1,\alpha}e^{-iS(a)}\int_0^\infty\frac{\sin^2(\frac{vS'}{2})}{v^{\alpha+1}}dv=|S'|^\alpha \Psi(a),
\end{equation}
where $S'=dS/da$.
Therefore, in the classical regime, the fractional
WDW equation yields the following fractional Hamiltonian constraint
\begin{equation}
    \label{sh7}
    -\frac{2\pi}{3{\mathcal V_0}m_P^\alpha a}|\Pi_a|^\alpha+
    \frac{3 {\mathcal V}_{0} m_\text{P}^{2}}{8 \pi L^2}a^3=0.
\end{equation}
The corresponding fractional ADM Hamiltonian, counterpart to (\ref{Hamil1}),
is
\begin{equation}
    \label{sh8}
    H_\text{ADM}^{(\alpha)}=N\left[ -\frac{2\pi}{3{\mathcal V_0}m_P^\alpha a}|\Pi_a|^\alpha
    + \frac{3 {\mathcal V}_{0} m_\text{P}^{2}}{8 \pi L^2}a^3\right].
\end{equation}
Hence, the WKB fractional wavefunction of the fractional
WDW Equation (\ref{FWDW}), for expanding universe, disregarding the preexponential factor, is
\begin{equation}
    \label{sh9}
    \Psi(a)\propto\exp{\left\{i\left( \frac{3\mathcal V_0m_P^{\frac{\alpha}{2}+1}}{4\pi L}\right)^\frac{2}{\alpha}a^\frac{4+\alpha}{\alpha}\right \}}.
\end{equation}
The Hamilton's equations,
\begin{equation}
    \label{sh10}
    \dot a=\frac{\partial H^{(\alpha)}_\text{ADM}}{\partial \Pi_a},~~~ \dot \Pi_a=-\frac{\partial H^{(\alpha)}_\text{ADM}}{\partial a},
\end{equation}
together with the Hamiltonian constraint (\ref{sh7}), in the comoving frame, $N=1$, lead to the following fractional solution:
\begin{equation}
    \label{sh11}
    a(t)=\left[\frac{2(D-2)}{(D-1)} \right]^\frac{1}{2(D-2)}\left(\frac{4\pi L}{3\mathcal V_0m_P}\right)^\frac{1}{2}\left(\frac{t}{L}\right)^\frac{1}{2(D-2)},~~~~~2<D<3,
\end{equation}
where according to Refs.~\cite{Jalalzadeh:2021gtq,Jalalzadeh:2022uhl}, we defined $D\equiv2/\alpha+1$.

In what follows, let us share some fascinating results for the fractional evolution of the universe during the fractional inflation epoch.

    Inflation is frequently achieved by having a scalar field with a suitable potential that dominates the early energy density of the universe. As a result of this energy density, the scale factor of the universe expands rapidly. In the original theories of inflation,  this expansion would be exponential and would resemble de Sitter space as the scalar field  steadily rolls down to its global minimum. Therefore, during the inflation, the deceleration parameter is $q=-1$. However, in fractional modification of the same model, we obtain
    \begin{equation}
        \label{sh12}
        q=-\frac{\ddot a/a}{H^2}=2D-5,
    \end{equation}
    where we used (\ref{sh11}).
    According to Equation \eqref{sh12}, the early universe is accelerated if $2<D<2.5$, and for these values of $D$, we have a power law inflationary model. Hence, the L\'evy’s fractional parameter $\alpha$ plays a crucial role in this model.\\

    Assuming that the inflationary epoch occurred during a time interval $t_f-t_i=10^{-37}~s=10^8t_P$ after the Planck's time, we obtain the number of e-foldings as
    \begin{equation}
        \label{sh13}
        N_e=\ln\left(\frac{a(t_f)}{a(t_i)}\right)=\frac{1}{2(D-2)}\ln\left(\frac{t_f}{t_i}\right)=\frac{\ln(10)}{2(D-2)}\simeq\frac{1}{D-2},
    \end{equation}
where we used the fractional form of scale factor (\ref{sh11}).
For instance, assuming $D=2.019$ we obtain $N_e\simeq60$, that is the minimum number of e-foldings necessary for resolving the standard problems of the Big Bang cosmology.


\section{Conclusions {and Discussions}}

Our simple fractional quantum cosmological model with characteristic features effectively influenced the early inflationary phase of the universe. Observations of CMB fluctuations and the distribution of matter in the universe constrain the properties of the inflaton field. A heavy scalar field with a mass of $10^{13}$ GeV, close to the grand unified theory scale, is preferred by well-known arguments even though the mass of such an inflaton and its interaction with matter fields are unknown. This field is widely cited as evidence that there are new physics at the intersection of the electroweak and Planck scales.
Our results imply that power-law inflation within fractional quantum cosmology may emerge from a new angle by accounting for quantum gravity effects via the fractional derivative.

In this paper, by considering a FLRW universe with all three curvature indices, a canonical kinetic energy associated with a scalar field minimally coupled to gravity, and a scalar potential, we obtained the corresponding Hamiltonian. Then, considering a non-zero ordering parameter, we constructed the standard WDW equation. To explore the goals of our paper, we focused on studying the early state of the universe. With this in mind, we confined ourselves to a semi-classical case, where the slow roll regime and a vanishing ordering parameter have been assumed for a compact flat universe, which represents the simplest case study to the problem of initial conditions~\cite{Linde:2004nz}.

The above assumptions motivated us to investigate our main objectives in two different contexts as follows.

\begin{description}
  \item[Standard quantum cosmology:]
 {For this case, the subsequent steps have been followed. }

\begin{itemize}
  \item  {According to the arguments presented above, we applied the slow roll regime, and substituted $p=0$ and $\mathcal K=0$ into generalized WDW Equation \eqref{wdw1}.}

  \item {We then showed that the solution of the simplified differential Equation \eqref{wdw1a} can be expressed by the Bessel function.}

  \item The model was then examined according to the arguments set out previously; please see the expressions below Equation \eqref{sh1}. Concretely, the model was studied assuming the WKB condition. For such a semiclassical regime, we have shown that the scale factor of the universe evolves exponentially at early times, which corresponds to the accelerated de Sitter~expansion.  

  \item We have additionally shown that $L\equiv\sqrt{3/\Lambda}$ ought to  take a value $10^7 l_p$ to obtain a sufficient e-folding number required to solve the problems of the standard cosmology.
\end{itemize}

 \item[Fractional quantum cosmology:] {The fractional extensions of the above-referred outcomes have been the principle goal of this paper. In what follows, allow us to summarize the relevant consequences and analyze them.  }
 \begin{itemize}
  \item {We gave a brief overview of fractional quantum mechanics and then explained how a suitable model is constructed for the fractional quantum cosmology.} {Based on the FQC procedure, we derived the fractional WDW equation and fractional ADM Hamiltonian from those retrieved for the corresponding standard model.}

  \item To solve the fractional WDW equation, we applied the WKB approximation again and set $N=1$.

  \item Despite the standard case, we have shown that the scale factor of the universe takes the form of a power-law function of time. {It is seen that it not only depends on the L\'evy’s fractional parameter $\alpha$, but also, as can be seen from the relation~(\ref{sh11}), on the volume of the compact 3-space $\Sigma$, $\mathcal V_0$. }
    \end{itemize}
    \end{description}

     {Based on the points summarized above, let us reiterate the main differences in the consequences of the standard quantum cosmology and FQC for our simple model.}

    \begin{enumerate}
 \item It is worth mentioning that the standard model and its fractional counterpart yield completely different results for the evolution of the scale factor: for the standard case, not only does the scale factor $a(t)$ accelerate with the exponential-law of the time, but also such evolution is only affected by matter (i.e., the constant potential; cf. Equation~\eqref{sh00}), and therefore the number of e-folding depends only on the constant $L$. In the fractional case, but the scale factor is a power-law function of time, which is completely independent of the matter; but instead, only the L\'evy’s fractional parameter $\alpha$ and the volume of the compact 3-space $\Sigma$, $\mathcal V_0$, determine how the scale factor evolves (cf. relation \eqref{sh11}).
  This gives us another dependency for the number of e-folding. Such consequences are interpreted as distinguishing features of the fractional quantum gravity.

 \item As mentioned, unlike the usual de Sitter solution (\ref{sh00}), the evolution of the scale factor associated with the fractional case, i.e., (\ref{sh11}), depends on the volume of the compact 3-space $\Sigma$, $\mathcal V_0$. More concretely, the evolution of the scale factor depends on the global geometry that affects the topology of the universe as a whole. The global geometry of the universe, i.e., the spatial curvature and topology and consequently the shape of the entire universe, are not determined by the Einstein gravitational field equations since they are differential equations that only determine the local features of space-time. Therefore, the 3-volume dependence of the scale factor is entirely a fractional quantum gravity effect. As we see in (\ref{sh11}), the value of the scale factor at early time $t=L$ is proportional to the inverse root square of $\mathcal V_0$, such that the smaller the value of the 3-volume, the bigger the initial scale factor.

Interestingly, the authors of Ref.~\cite{Aurich:2007yx} (for a recent analysis, see~\cite{Aslanyan:2013lsa,Aurich:2008km,Planck:2013okc}) have shown that the WMAP 3 year data are 
consistent with the possibility that we live in a `small universe' shaped like a flat
3-torus $\mathbb T^3$ whose fundamental domain is a cube with side length $l =4$ corresponding to a volume of $5 \times 10^3$ Gpc$^3$.
It turns out that the
torus model describes the data much better than the best-fit $\Lambda$CDM model, as it shows  the
suppression of the CMB anisotropy at large scales first observed by COBE.

\item As seen in Equation (\ref{sh5}), the Riesz fractional derivative is a non-local operator that can explain non-local processes in the minisuperspace. In fact, non-locality is a general behavior of all fractional derivatives and integrals and hence they describe processes with non-locality in time (memory) and space (large jumps)~\cite{doi:10.1142/11107}. The fractional derivative has lagged far behind the integer-order calculus due to the unclear physical meaning, which is a major obstacle. In 1974, it was proposed as an open problem to ask ``What are the physical interpretations of fractional calculus?'' Using analogous reasoning, a physical explanation for a fractional time derivative was put forth in 2002~\cite{2001mathP}, although no experiments were performed to support the new time scale. In Ref.~\cite{2013NatSR431D}, the authors have shown that the time fractional order is a memory index. Time disappears in quantum gravity and cosmology, and the space-fractional derivative may play a crucial role. In Refs.~\cite{Jalalzadeh:2021gtq,Jalalzadeh:2022uhl}, the authors show that L\'evy's fractional parameter $\alpha$ represents the fractal dimension (denoted by $D$ in Equations~(\ref{sh11}), (\ref{sh12}) and (\ref{sh13})) of the black hole horizon or the cosmological horizon. In addition, as we saw in (\ref{sh11}), as a result of big jumps in minisuperspace, the initial value of emerged classical scale factor depends on the global geometry (or topology) of the universe. Furthermore, Equation (\ref{sh13}) shows that the acceleration of the universe and e-folding of the inflation epoch are direct consequences of its topology.

   \end{enumerate}

\vspace{6pt}

\section{ACKNOWLEDGMENTS}
S.M.M.R. and P.M. acknowledge the FCT grants UID-B-MAT/00212/2020 and UID-P-MAT/00212/2020
at CMA-UBI plus the COST Action CA18108 (Quantum gravity phenomenology in the multi-messenger approach).

\end{document}